%% file: paper.tex
\pdfoutput=1                           

\documentclass{article}

\usepackage[T1]{fontenc}                
\usepackage{microtype}
\usepackage{graphicx}
\graphicspath{{figures/}}              
\usepackage{subcaption}
\usepackage{booktabs}
\usepackage{adjustbox}                 
\usepackage{amsmath}
\usepackage{amssymb}
\usepackage{mathtools}
\usepackage{xcolor}
\usepackage[most]{tcolorbox}
\usepackage{enumitem}                  
\usepackage{pifont}                    
\usepackage{tikz}                      
\usetikzlibrary{arrows.meta,positioning}

\definecolor{passgreen}{RGB}{20,120,60}
\definecolor{failred}{RGB}{190,40,40}
\newcommand{\pass}{\textcolor{passgreen}{\ding{51}\,pass}}
\newcommand{\fail}{\textcolor{failred}{\ding{55}\,\textbf{fail}}}

\newtcolorbox{takeaway}{
  colback=blue!4, colframe=blue!35, boxrule=0.5pt,
  left=5pt, right=5pt, top=3pt, bottom=3pt, arc=2pt}

\usepackage{hyperref}

\usepackage[preprint]{icml2026}

\usepackage[capitalize,noabbrev]{cleveref}

\newtcolorbox{promptbox}{
  colback=gray!5, colframe=gray!55, boxrule=0.4pt,
  left=4pt, right=4pt, top=4pt, bottom=4pt,
  fontupper=\small\ttfamily, breakable
}

\icmltitlerunning{Instruction Stacking Collapse}

\begin{document}

\twocolumn[
  \icmltitle{Instruction Stacking Collapse: A Benchmark and the\\
             Capability-Dependent Value of Prompt Compilation}

  \begin{icmlauthorlist}
    \icmlauthor{Atul Anand}{}
    \icmlauthor{Sourav Chattaraj}{}
  \end{icmlauthorlist}

  \icmlkeywords{instruction following, large language models, prompt engineering,
                benchmark, constraint satisfaction, prompt compilation, evaluation}
  \vskip 0.3in
]

\makeatletter\icml@noticeprintedtrue\makeatother

\input{abstract}

\input{body}

\section*{Impact Statement}
This work studies the reliability of instruction-following in deployed language
models and a cheap mitigation for it. The likely effect is positive: better
adherence to safety-, format-, and policy-constraints on the small models that
serve most production traffic. Two cautions apply. First, our benchmark measures
\emph{compliance}, not the \emph{correctness or safety} of content, so higher
follow rates must not be read as higher trustworthiness. Second, the
capability-graded finding could be misread as ``frontier models need no prompt
hygiene''; our data show only that one specific restructuring helps them little,
not that they are immune to instruction collapse. No human subjects or private
data are involved, and all artifacts are released to support scrutiny.

\section*{Software and Data}
All code, the 24-instruction benchmark, deterministic verifiers, cached model
responses, per-row result parquets, and the analysis scripts (including the
cluster-robust re-analysis of \cref{sec:compiler}) are released to enable exact
reproduction without new API calls.

\bibliography{references}
\bibliographystyle{icml2026}

\newpage
\appendix
\onecolumn
\input{appendix}

\end{document}

%% file: abstract.tex
\begin{abstract}
Production prompts rarely carry a single instruction. One system message may require
valid JSON, a word limit, three citations, and a fixed tone at the same time. We
study how instruction-following degrades as such constraints accumulate. We introduce
a benchmark that stacks 24 verifier-checked instructions, one to twenty at a time, and
evaluate three production-tier LLMs (Claude Sonnet~4.6, GPT-5-mini, Gemini~2.5~Flash).
Instruction-following degrades non-linearly: the follow rate falls from ${\sim}96\%$
to as low as $20\%$, driven by a structured and reproducible set of pairwise
conflicts. A single ``output JSON'' constraint, for example, is jointly unsatisfiable
with nine others. We then evaluate a training-free remedy: an \emph{instruction
compiler} that rewrites the stacked prompt in a single LLM call and is reused across
queries. Its benefit is \emph{capability-graded}. It recovers up to ${+}11$ points of
follow rate for weaker models, which are also the models most often deployed at scale,
while leaving stronger models, which already internalise the same structure,
essentially unchanged. Cluster-robust tests, same-baseline controls, and a
within-family scaling ladder attribute the gain to the rewrite itself rather than to
additional tokens, reordering, or measurement headroom. We release the benchmark,
verifiers, and cached runs for full reproduction.
\end{abstract}

%% file: body.tex
\section{Introduction}
\label{sec:intro}

Large language models reliably follow a single instruction, but production prompts
rarely contain only one. A deployed system prompt may at once require a specific
output format, a length limit, citations, and a fixed tone. When many such
constraints are combined, models satisfy fewer of them, and the violations are
silent: no error is raised when an instruction is dropped. It is widely reported that
compliance falls as the number of constraints grows, yet there is no systematic
account of how quickly it falls, which instructions interfere with one another, or
whether the loss can be recovered at inference time.

\cref{tab:motiv} makes the failure concrete. Given eight stacked instructions,
GPT-5-mini obeyed ``output JSON'', and that one choice silently broke four others.
No error was raised; four constraints simply went unmet. \cref{sec:conflict} shows
this is the rule, not the exception.

\begin{table}[t]
\centering
\caption{One response, eight stacked instructions, four silent failures. Satisfying
F1 (JSON) made four markdown-/case-dependent instructions impossible, a cascade from
a single structural choice (GPT-5-mini, advice task).}
\label{tab:motiv}
\begin{adjustbox}{max width=\columnwidth}
\begin{tabular}{lll}
\toprule
Instruction (8 stacked) & Verdict & Why \\
\midrule
C1 reference a year    & \pass & matched ``2021'' \\
F1 valid JSON          & \pass & parses \\
L1 $\le$50 words       & \pass & 29 words \\
X1 use ``therefore''   & \pass & ${\times}1$ \\
F4 ``Summary:'' line   & \fail & JSON has no bare trailing line \\
L4 title $\le$10 words & \fail & first line is the 29-word JSON \\
R2 \#\#Assumptions     & \fail & no markdown header inside JSON \\
X4 all-lowercase       & \fail & JSON keys force an uppercase char \\
\bottomrule
\end{tabular}
\end{adjustbox}
\end{table}

Existing benchmarks do not chart this. IFEval~\citep{zhou2023ifeval} measures one or
two instructions per prompt; FollowBench~\citep{jiang2024followbench} stacks
constraints but only within a single category. Both answer ``can the model obey
instruction $X$?'' Neither answers ``what happens when a dozen instructions from
different categories must all hold at once?'' That is the question deployment
actually poses.

\textbf{This paper maps the collapse and tests a fix.} We build a benchmark of 24
atomic instructions across six categories (format, length, lexical, structural,
reasoning, content), each paired with a released verifier, and stack them uniformly
at random into prompts of size 1--20. We evaluate three production-tier models, one
per major provider. We then ask three questions and, because we pre-registered our
predictions, report plainly where the data confirmed them and where it did not.

\begin{figure*}[t]
\centering
\begin{tikzpicture}[font=\footnotesize, >=Latex, node distance=8mm and 9mm,
  box/.style={rounded corners=2pt,draw=gray!60,fill=gray!6,align=center,inner sep=4pt,text width=27mm,minimum height=11mm},
  llm/.style={rounded corners=2pt,draw=gray!70,fill=gray!12,align=center,inner sep=3pt,text width=9mm,minimum height=11mm},
  proc/.style={rounded corners=2pt,draw=blue!55,fill=blue!7,align=center,inner sep=4pt,text width=32mm,minimum height=11mm},
  bad/.style={rounded corners=2pt,draw=failred!75,fill=failred!7,align=center,inner sep=4pt,text width=27mm,minimum height=11mm},
  good/.style={rounded corners=2pt,draw=passgreen!75,fill=passgreen!8,align=center,inner sep=4pt,text width=33mm,minimum height=11mm},
  arr/.style={->,thick,gray!65}]
  \node[box] (prompt) {Stacked prompt\\\textbf{1--20 instructions}\\{\scriptsize JSON\,$\cdot$\,$\le$50w\,$\cdot$\,5 bullets\,$\cdot$\,\#\#Assumptions\,$\cdots$}};
  \node[llm,right=of prompt] (llm1) {LLM};
  \node[bad,right=of llm1] (collapse) {\textbf{Collapse} (\cref{sec:collapse})\\follows 20--60\%\\at 20 instructions};
  \node[proc,below=9mm of prompt] (comp) {\textbf{Instruction compiler}\\{\scriptsize one LLM call:}\\{\scriptsize cluster\,$\cdot$\,merge\,$\cdot$\,precedence}};
  \node[llm,right=of comp] (llm2) {LLM};
  \node[good,right=of llm2] (rec) {\textbf{Capability-graded recovery} (\cref{sec:compiler})\\weak $+11$\,pp\,$\cdot$\,mid $+3$\,pp\\strong $-1$\,pp};
  \draw[arr] (prompt)--(llm1);
  \draw[arr] (llm1)--(collapse);
  \draw[arr] (prompt.south)--(comp.north);
  \draw[arr] (comp)--(llm2);
  \draw[arr] (llm2)--(rec);
\end{tikzpicture}
\caption{\textbf{Overview.} Stacking many instructions into one prompt collapses
instruction-following (top). A training-free \emph{instruction compiler} rewrites the
stack in one LLM call; its benefit is \emph{capability-graded}: large for weak
models, neutral-to-negative for strong ones (bottom).}
\label{fig:overview}
\end{figure*}
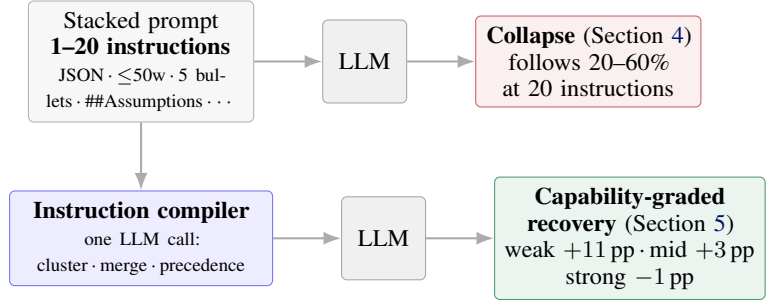

\textbf{Collapse (RQ1).} Follow rate decays smoothly but non-linearly with stack
size, from ${\sim}0.96$ at one instruction to $0.60$ (Sonnet), $0.43$ (Gemini), and
$0.20$ (GPT-5-mini) at twenty. A sigmoid (flat near baseline, then a bend) fits
two of three models, and a bootstrap over the model-selection itself picks it in
93--100\% of refits, so the shape is not an artifact of fitting four parameters to
seven points.

\textbf{Conflict (RQ2).} The cause is interference between instructions. Testing all 231
pairs and separating the 15 that are \emph{logically impossible} (valid JSON cannot
also carry markdown headers) from genuine \emph{behavioral} failures, we find
${\sim}12\%$ of the satisfiable pairs (95\% CI $[4.6, 11.6]$) still fail far more
than their individual rates predict. The structure is interpretable and
reproducible: ``output JSON'' is the dominant conflict hub, ``cite named sources''
the dominant synergy, and the whole interaction landscape correlates across both
tasks and all three models ($\rho = {+}0.23$ to ${+}0.28$).

\textbf{Mitigation (RQ3).} We then ask whether the loss can be recovered without
training. We evaluate an \emph{instruction compiler}: a single LLM call that rewrites
the raw stack by clustering instructions by category, merging redundant ones, and
inserting a precedence note where two rules conflict. The rewrite is computed once
per instruction set and reused across queries. Its effect is \emph{capability-graded}.
It recovers ${+}11.0$~pp of follow rate for the weakest target (GPT-5-mini) and
${+}3.3$~pp for the middle one (Gemini), while leaving the strongest target (Sonnet)
essentially unchanged (${-}1.2$~pp, not statistically robust). The recovery is thus
concentrated on the smaller models that are most often deployed at scale, where the
need is greatest. The effect is smaller than the $d \ge 0.5$ we pre-registered, but
its source is well identified: three controls attribute it to the semantic rewrite
rather than to additional tokens, reordering, or the compiler grading its own output,
and a within-family scaling ladder reproduces the ordering (Spearman ${-}0.85$
between raw strength and recovery).

\textbf{Interpretation.} Two readings are consistent with this pattern. The
\emph{practical} reading: because the rewrite is computed once and reused, it is an
inexpensive intervention precisely for the smaller, faster models that serve most
production traffic, and is unnecessary for top-tier models. The \emph{mechanistic}
reading: a strong model reconstructs the stack's structure on its own while reading
the raw prompt, whereas a weaker model does not, so making that structure explicit
helps the weaker model and is redundant for the stronger one. We do not overstate the
magnitude: even in the best case the weakest model still leaves about two-thirds of
its instructions unsatisfied at the largest stack size.

\textbf{Contributions.}
\begin{enumerate}[topsep=2pt,itemsep=1pt,leftmargin=1.4em]
  \item A \textbf{stacked-instruction benchmark} (24 verifier-backed instructions,
        six categories, random stacks of size 1--20) and an audited verifier suite
        (\cref{sec:benchmark}).
  \item A \textbf{characterization of collapse} across three models and three task
        types, with bootstrap-stable functional forms and the finding that collapse
        is universal but per-model ordering is task-dependent
        (\cref{sec:collapse,sec:multitask}).
  \item A \textbf{conflict topology} that separates designed-in impossibilities from
        behavioral conflicts and is shared across tasks and models
        (\cref{sec:conflict}).
  \item A \textbf{training-free instruction compiler} and a controlled evaluation of
        it: a capability-graded recovery that is largest for weaker models, with the
        active mechanism isolated by ablation and a within-family ladder that removes
        the cross-lab confound (\cref{sec:compiler}). To our knowledge this is the
        first analysis of how the value of a prompt rewrite depends on target
        capability.
\end{enumerate}
All code, cached responses, verifiers, and analysis scripts are released.

\section{Related Work}
\label{sec:related}

\textbf{Instruction-following benchmarks.} IFEval~\citep{zhou2023ifeval} pioneered
regex-verifiable instructions but places only one or two per prompt, so it cannot
see degradation with count. FollowBench~\citep{jiang2024followbench} stacks
constraints \emph{within} a single category; ours is the random cross-category
analog, exposing the cross-category interactions (\cref{sec:conflict}) a
within-category design cannot. RECAST~\citep{guo2025recast} and
MulDimIF~\citep{ye2026muldimif} push multi-constraint \emph{training data}; we target
\emph{evaluation} and \emph{mitigation} (\cref{tab:related}; full version in
\cref{app:related}).

\begin{table}[t]
\centering
\caption{Benchmark comparison (full version in \cref{app:related}).}
\label{tab:related}
\begin{adjustbox}{max width=\columnwidth}
\begin{tabular}{lccccc}
\toprule
Property & IFEval & FollowB. & RECAST & MulDim & \textbf{Ours} \\
\midrule
Instr./prompt        & 1--2 & 1 cat & 30+ & multi & \textbf{1--20} \\
Cross-category       & no & no & part. & part. & \textbf{yes} \\
Pairwise conflict    & no & no & data & data & \textbf{FDR} \\
Imposs.-vs-behav.    & no & no & no & no & \textbf{yes} \\
Train-free mitig.    & no & no & no & no & \textbf{yes} \\
\bottomrule
\end{tabular}
\end{adjustbox}
\end{table}

\textbf{Prompt rewriting and inference-time mitigation.} Several methods improve
constraint-following by editing the prompt at inference time.
PRewrite~\citep{kong2024prewrite} trains a prompt rewriter with reinforcement
learning; DeCRIM~\citep{ferraz2024decrim} decomposes an instruction into atomic
constraints and runs an iterative critique-and-refine loop; RECAST~\citep{guo2025recast}
and MulDimIF~\citep{ye2026muldimif} instead curate multi-constraint \emph{training}
data. All of these require either model training or several model calls per query. Our
instruction compiler differs in being a single, zero-shot, training-free rewrite that
is computed once per instruction set and reused across all queries, adding no
per-query cost. More importantly, prior work treats prompt rewriting as a uniform
improvement; we are not aware of any study that characterises how its value varies
with the capability of the target model. That dependence, rather than the rewrite
itself, is the methodological contribution of \cref{sec:compiler}.

\textbf{Capability-dependent prompt engineering.} Our central claim, that one compilation
step helps the \emph{weak} model far more than the strong one, is the mirror image
of \citet{hakim2026brevity}, where brevity constraints help the \emph{strong} model
by curbing over-elaboration. Both say a prompt-level fix has model-dependent value.
This is not inverse scaling on the task~\citep{mckenzie2023inverse}, since the task is
fixed, but inverse scaling on the \emph{value of a prompt-engineering move}.
\citet{qi2026paradox} note a related paradox where models fail by adding a
self-evident constraint already met implicitly.

\section{The Stacking-Collapse Benchmark}
\label{sec:benchmark}

\textbf{Instructions.} 24 atomic instructions, four per category
(\cref{tab:categories}). Each has one canonical phrasing shown verbatim and a
verifier mapping a response to pass/fail. Two (analogy, counterargument) need an LLM
judge and are excluded, leaving \textbf{22 deterministic instructions} as the active
pool; the full list is in \cref{app:instructions}.

\begin{table}[t]
\centering
\caption{The six categories (representative instructions; full set in
\cref{app:instructions}).}
\label{tab:categories}
\begin{adjustbox}{max width=\columnwidth}
\begin{tabular}{ll}
\toprule
Category & Examples \\
\midrule
Format (F1--4)     & valid JSON; code-fence; $\ge$3 bullets; ``Summary:'' line \\
Length (L1--4)     & $\le$50 words; $\ge$150 words; exactly 5 bullets; title $\le$10 \\
Lexical (X1--4)    & use ``therefore''; avoid ``however''; 3 citations; lowercase \\
Structural (S1--4) & begin ``Answer:''; \#\#Pros/\#\#Cons; 2nd-person; no ``?'' \\
Reasoning (R1--4)  & \#\#Reasoning; \#\#Assumptions; Confidence:0.XX; \#\#Alt. \\
Content (C1--4)    & a year; analogy$^*$; counterarg$^*$; $\ge$2 sources \\
\bottomrule
\end{tabular}
\end{adjustbox}
\\[2pt]
{\footnotesize $^*$require an LLM judge; excluded from the active pool.}
\end{table}

\textbf{Verifiers.} Each verifier is \emph{strict on intent, lenient on cosmetic
markup} (case-insensitive headers; any list marker counts; an optional code-fence
around JSON is stripped). These policies came from a pilot audit that flipped 8.0\%
of verdicts. A human audit (${\sim}500$ labels over 19 verifiers) finds
\textbf{18/19 agree with humans at 100\%}, one (C4) at 80\%, and caught and fixed a
real bug in L4 that was re-scored from cache with no new calls. Three verifiers (R1,
R3, S2) are spot-checked only; a full 100-label-per-instruction validation is owed
(\cref{sec:discussion}). Details and the agreement table are in \cref{app:verifiers}.

\textbf{Grid.} Stacks are sampled \textbf{uniformly at random} (not screened for
feasibility, since real prompts are not), at sizes $\{1,3,5,8,12,16,20\}$, 30 stacks/size
$\times$ 10 items. Three models: \textbf{Claude Sonnet 4.6}, \textbf{GPT-5-mini},
\textbf{Gemini 2.5 Flash}, temperature 0. The primary task is open-ended
\emph{advice}; GSM8K \emph{math} and HumanEval \emph{code} are generalization checks
(\cref{sec:multitask}). All responses are cached by $(\text{model},
\text{system}, \text{user})$ hash for exact reproducibility.

\section{Characterizing the Collapse}
\label{sec:collapse}

\subsection{Degradation is real and non-linear}
\label{sec:degradation}

\begin{figure}[t]
\centering
\includegraphics[width=\columnwidth]{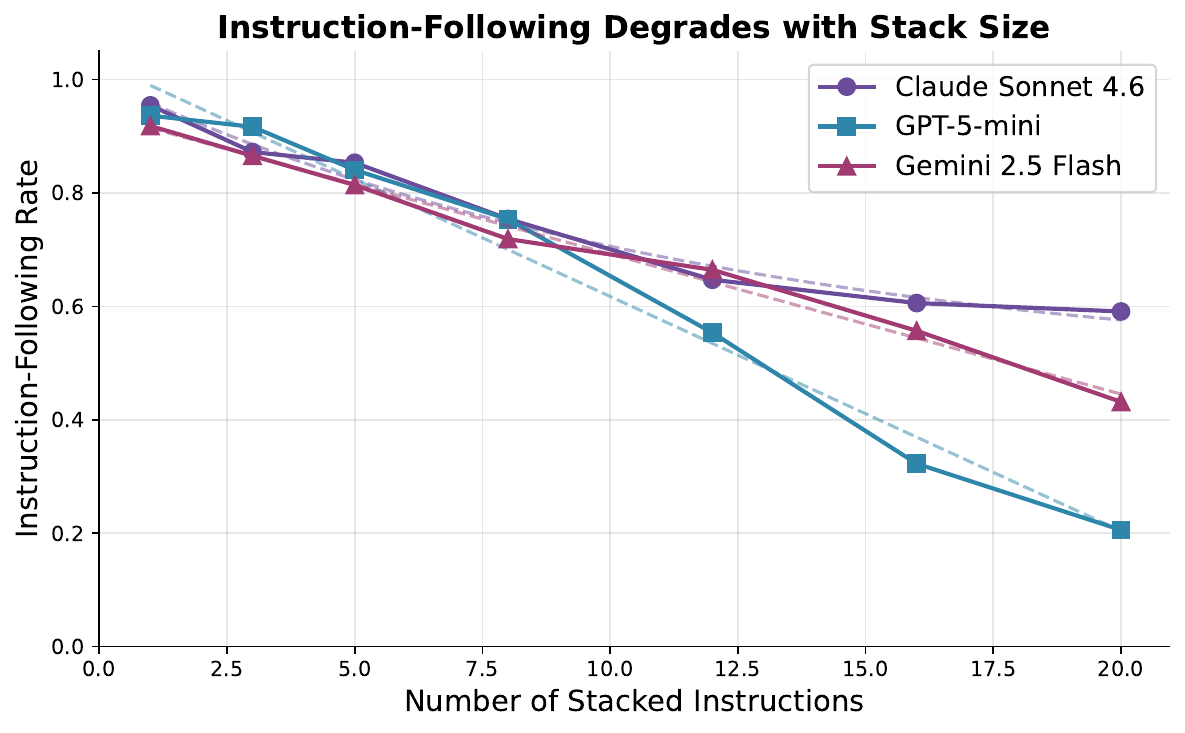}
\caption{Instruction-following rate vs.\ stack size (advice). Each point is a mean
over 30 random stacks $\times$ 10 items (${\sim}300$ trials). Follow rate falls from
${\sim}0.96$ to $0.20$--$0.60$; GPT-5-mini collapses sharpest.}
\label{fig:degradation}
\end{figure}

Follow rate falls from ${\sim}0.96$ at one instruction to $0.604$ (Sonnet), $0.433$
(Gemini), and $0.201$ (GPT-5-mini) at twenty (\cref{tab:rates}, \cref{fig:degradation}).
Cluster-bootstrap CIs (resampling responses) are mutually non-overlapping at stack
$\ge 16$, so the ordering is reliable. The shape is non-linear: an AIC comparison of
linear / exponential / sigmoid selects \textbf{sigmoid} for Sonnet and GPT-5-mini,
and a bootstrap over the model selection itself confirms it is stable (sigmoid wins
\textbf{93\%} of refits for Sonnet, \textbf{100\%} for GPT-5-mini). Gemini is
\textbf{non-sigmoidal}: its decline begins immediately and is best described as
linear-to-exponential (sigmoid in $<2\%$ of refits). Categories degrade unevenly:
lexical survives best, format and length collapse hardest (\cref{app:categories}).

\begin{table}[t]
\centering
\caption{Mean follow rate, cluster-bootstrap 95\% CI (advice).}
\label{tab:rates}
\begin{adjustbox}{max width=\columnwidth}
\begin{tabular}{lcccc}
\toprule
Model & stack=1 & stack=5 & stack=16 & stack=20 \\
\midrule
Sonnet 4.6 & .964\,[.94,.99] & .827\,[.81,.85] & .621\,[.60,.64] & .604\,[.59,.62] \\
GPT-5-mini & .964\,[.94,.99] & .845\,[.82,.87] & .259\,[.24,.28] & .200\,[.19,.21] \\
Gemini 2.5 F & .959\,[.93,.99] & .823\,[.80,.84] & .535\,[.52,.56] & .433\,[.41,.45] \\
\bottomrule
\end{tabular}
\end{adjustbox}
\end{table}

We operationalize \textbf{target strength} as a model's raw stacked-following rate
(Sonnet $>$ Gemini $>$ GPT-5-mini), the axis \cref{sec:compiler} grades the compiler
against.

\subsection{The failure is structured: a conflict topology}
\label{sec:conflict}

Recall \cref{tab:motiv}: one structural commitment, emitting JSON, silently broke
four other instructions in a single response. That cascade is not an accident of one
prompt; it is the aggregate pattern, which we now quantify across all pairs.

\begin{figure}[t]
\centering
\includegraphics[width=\columnwidth]{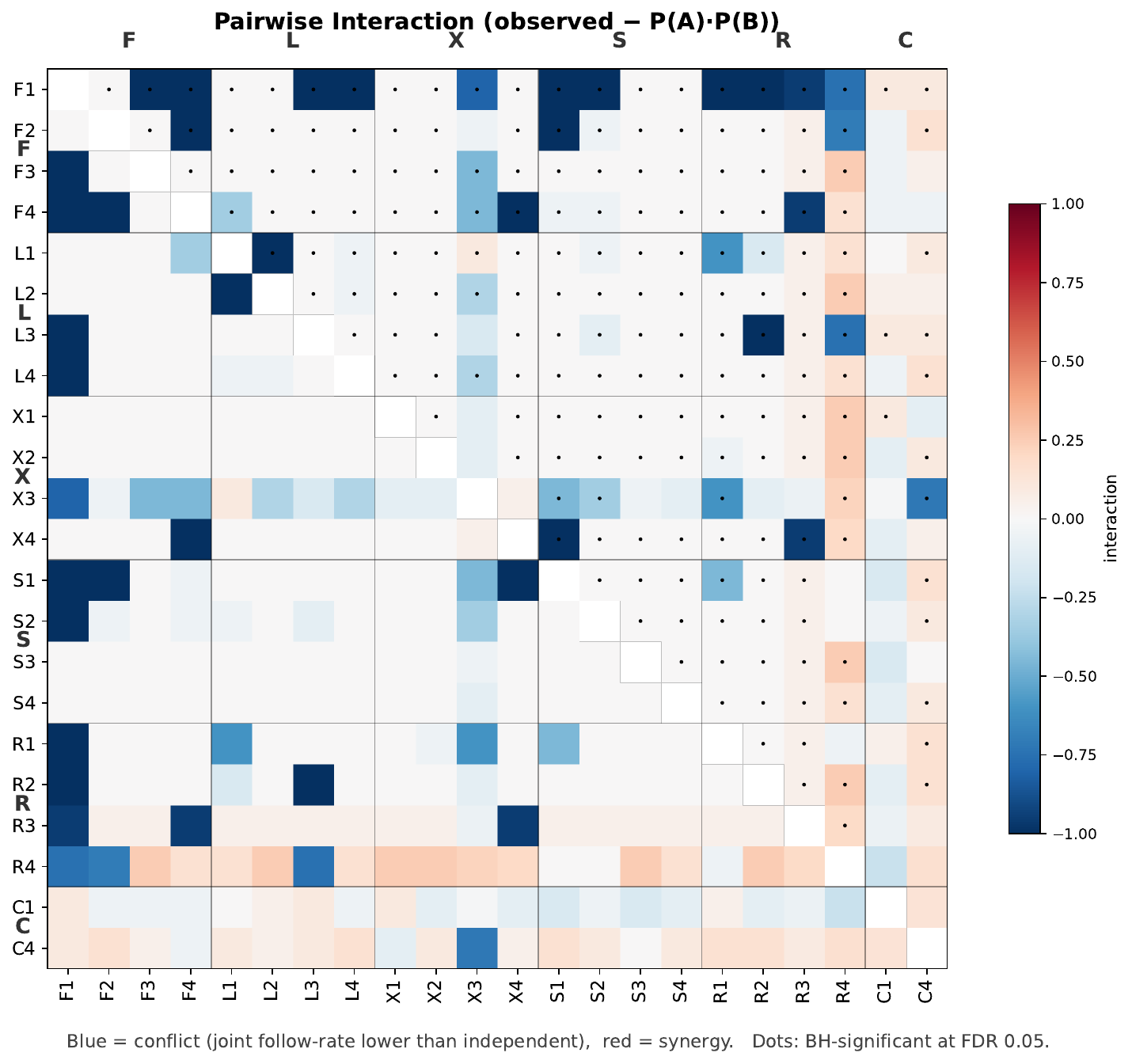}
\caption{Pairwise interaction for the 231 pairs (Sonnet), ordered by category.
Blue $=$ conflict (joint rate below independent baseline); red $=$ synergy; dots are
BH-significant at FDR 0.05. F1 (JSON) is the dominant conflict row; C4 (named
sources) anchors the synergies.}
\label{fig:conflict}
\end{figure}

Testing all 231 pairs (interaction $= \text{observed}(A\wedge B) - P(A)\,P(B)$,
bootstrap CI, BH-corrected at FDR 0.05; \cref{fig:conflict}), we separate two things a
raw count conflates:
\begin{itemize}[topsep=2pt,itemsep=1pt,leftmargin=1.2em]
  \item \textbf{15 logical impossibilities} (e.g.\ JSON $\times$ any markdown
        instruction; $\le$50 vs.\ $\ge$150 words) sit at the $-1.0$ floor by
        construction: the compositional ceiling, reported not hidden.
  \item \textbf{Behavioral conflicts:} of the 216 satisfiable pairs,
        \textbf{${\sim}12\%$ (95\% CI $[4.6, 11.6]$)} fail significantly more than
        their parts predict, that is, pairs the model \emph{could} satisfy but does not
        (e.g.\ ``exactly 5 bullets'' $\times$ ``Assumptions section'', joint rate
        $0.00$).
\end{itemize}
The structure is interpretable: \textbf{F1 (JSON) is the dominant conflict hub},
\textbf{C4 (named sources) the dominant synergy} (a required section gives citations
a home). Crucially it is \textbf{reproducible}: the interaction landscape correlates
across all three models (Spearman $\rho = {+}0.27, {+}0.28, {+}0.23$;
\cref{app:conflict}) and across tasks (advice$\leftrightarrow$math $\rho = {+}0.44$).
The exact conflict \emph{count} is sensitive to the significance rule; the
\emph{topology} (rank agreement) is not, so we lead with the latter.

\subsection{What generalizes across tasks (and what does not)}
\label{sec:multitask}

Re-running the curve on GSM8K math and HumanEval code (\cref{app:multitask}):
\textbf{collapse is universal}: every model loses heavily by stack=20 on all three
tasks. But two things are task-dependent. The \emph{curve shape} shifts from
sigmoidal (prose) to linear (code). And the \emph{per-model ordering} is \textbf{not}
task-invariant: GPT-5-mini, the worst follower on advice/math ($0.20$), nearly ties
Sonnet on code ($0.44$) and overtakes Gemini. ``Model $X$ is the weakest follower''
is therefore false in general; weakness is joint in in model and task.

\section{Instruction Compilation: Method and Results}
\label{sec:compiler}

\subsection{The compiler}

One LLM call rewrites the raw stack via three transforms: \textbf{(1)~cluster +
merge} by category; \textbf{(2)~conflict resolution}, inserting a precedence note
where two rules conflict; \textbf{(3)~structural reformat}, producing a numbered checklist,
reasoning first (nearest ``thinking''), format last (nearest token production). It is
\textbf{training-free} (any chat API), \textbf{amortized} (compiled once, reused
across queries), and \textbf{falsifiable} (pre-registered paired $d \ge 0.5$). The
full prompt:

\begin{promptbox}
System: You are an expert prompt engineer. Restructure instructions for maximum compliance.

User:
You will receive a list of instructions that must ALL be followed.
Rewrite them as a single optimized instruction set that maximizes the chance all constraints are satisfied simultaneously.
Apply these rules:
1. Group by type: REASONING -> CONTENT -> STRUCTURAL -> LEXICAL -> LENGTH -> FORMAT
2. Within each group, merge instructions that overlap into one clearer statement.
3. If two instructions conflict, add: "When X conflicts with Y, prioritize X."
4. Output as a numbered checklist with category headers.
5. End with: "Before finalizing: verify EVERY numbered constraint is satisfied."
Do NOT add or remove constraints. Only restructure and clarify.

Raw instructions:
\{instructions\}
\end{promptbox}

In all experiments the compiler is Claude Sonnet 4.6. We test four conditions at
stacks $\{8,12,16,20\}$: \texttt{raw}, \texttt{compiled}, \texttt{expanded\_raw}
(token-padded control), and \texttt{reorder\_only} (deterministic category-reorder,
no LLM). Targets span same-model (Sonnet$\to$Sonnet) and cross-model
(Sonnet$\to$GPT-5-mini, Sonnet$\to$Gemini).

\subsection{Recovery is capability-graded}
\label{sec:recovery}

The compiler produces a recovery that grows as the target model weakens
(\cref{tab:recovery}). We report it with \textbf{cluster-robust} inference (resampling
whole stacks, since the items under one compiled prompt are not independent and a
naive $n{=}1{,}200$ t-test overstates significance). The recovery is smaller than the
$d \ge 0.5$ effect we pre-registered (\cref{sec:scorecard}), but it is statistically
significant for the two weaker targets and is concentrated where the need is greatest.

\begin{figure}[t]
\centering
\includegraphics[width=\columnwidth]{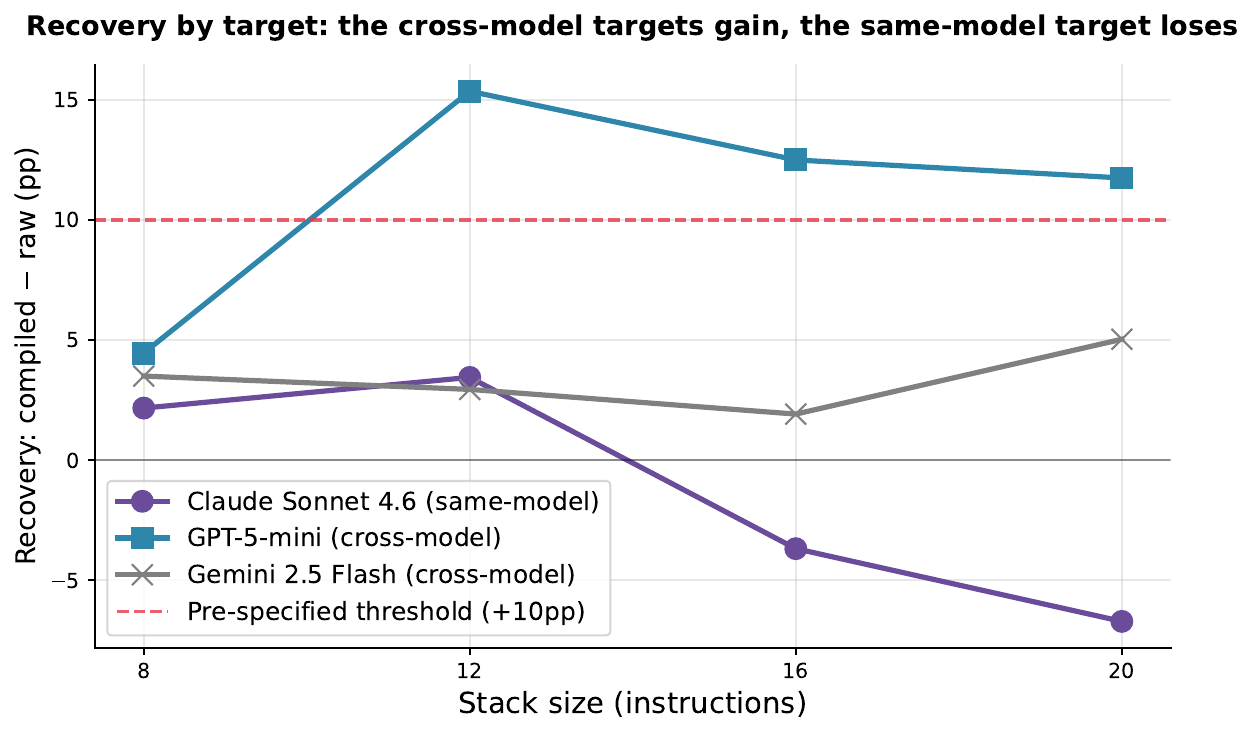}
\caption{Recovery (\texttt{compiled}$-$\texttt{raw}) by stack size per target;
dashed line is the pre-registered $+10$~pp threshold. GPT-5-mini clears it from
stack=12; Sonnet goes negative.}
\label{fig:recovery}
\end{figure}

\begin{table}[t]
\centering
\caption{Compiler recovery, pooled stacks 8--20. Cluster-bootstrap resamples whole
stacks; contrast with the naive t-test $p$.}
\label{tab:recovery}
\begin{adjustbox}{max width=\columnwidth}
\begin{tabular}{lcccc}
\toprule
Target & Mean (pp) & $d$ & Naive $p$ & \textbf{Cluster CI / $p$} \\
\midrule
Sonnet (same)    & $-1.2$  & $-0.06$ & $0.03$    & $[-4.7,+2.2]$ / $0.49$ \\
Gemini (cross)   & $+3.3$  & $+0.20$ & ${<}10^{-11}$ & $[+1.2,+5.5]$ / $0.001$ \\
GPT-5-mini (cross)& $+11.0$ & $+0.43$ & ${<}10^{-46}$ & $[+7.9,+14.5]$ / ${<}0.001$ \\
\bottomrule
\end{tabular}
\end{adjustbox}
\end{table}

The two cross-model gains are robust; the strong-model ``harm'' is \textbf{not}: null
when pooled and only marginal at the hardest stack size (cluster $p=0.037$ at
stack=20, stack-level sign test 14/20, $p=0.11$). So the accurate statement is:
compilation is \textbf{neutral for the strongest target and increasingly helpful as
the target weakens.} The ordering tracks target strength across providers.

\subsection{The mechanism is the LLM rewrite}
\label{sec:mechanism}

Four alternatives, all ruled out (\cref{tab:ablation}):
\begin{itemize}[topsep=2pt,itemsep=1pt,leftmargin=1.2em]
  \item \textbf{Not extra tokens / not reordering.} \texttt{expanded\_raw} and
        \texttt{reorder\_only} barely move the weak targets. Because they share the
        \emph{same baseline} as raw, ``compiled beats them'' cannot be a headroom
        artifact, yet compiled exceeds both for \emph{every} model (GPT-5-mini
        ${+}10.0/{+}12.1$~pp; Sonnet ${+}6.2/{+}4.1$~pp; all $p<10^{-24}$). This is
        the clustering-robust not-headroom argument.
  \item \textbf{Not leakage.} If the compiler merely fed Sonnet its own reasoning,
        the same-model case would gain most; instead it gains \emph{least}, while two
        \emph{different} models gain more.
  \item \textbf{Not a ceiling artifact.} The within-family ladder
        (\cref{sec:ladder}) shows the ordering where only capability varies.
\end{itemize}

\begin{table}[t]
\centering
\caption{Stack=20 follow rate by condition (advice).}
\label{tab:ablation}
\begin{adjustbox}{max width=\columnwidth}
\begin{tabular}{lccc}
\toprule
Condition & Sonnet & Gemini & GPT-5-mini \\
\midrule
\texttt{raw}          & 0.627 & 0.453 & 0.204 \\
\textbf{\texttt{compiled}} & \textbf{0.560} & \textbf{0.503} & \textbf{0.322} \\
\texttt{expanded\_raw}& 0.486 & 0.482 & 0.212 \\
\texttt{reorder\_only}& 0.501 & 0.463 & 0.214 \\
\bottomrule
\end{tabular}
\end{adjustbox}
\end{table}

\subsection{Within-family ladder}
\label{sec:ladder}

\begin{figure}[t]
\centering
\includegraphics[width=\columnwidth]{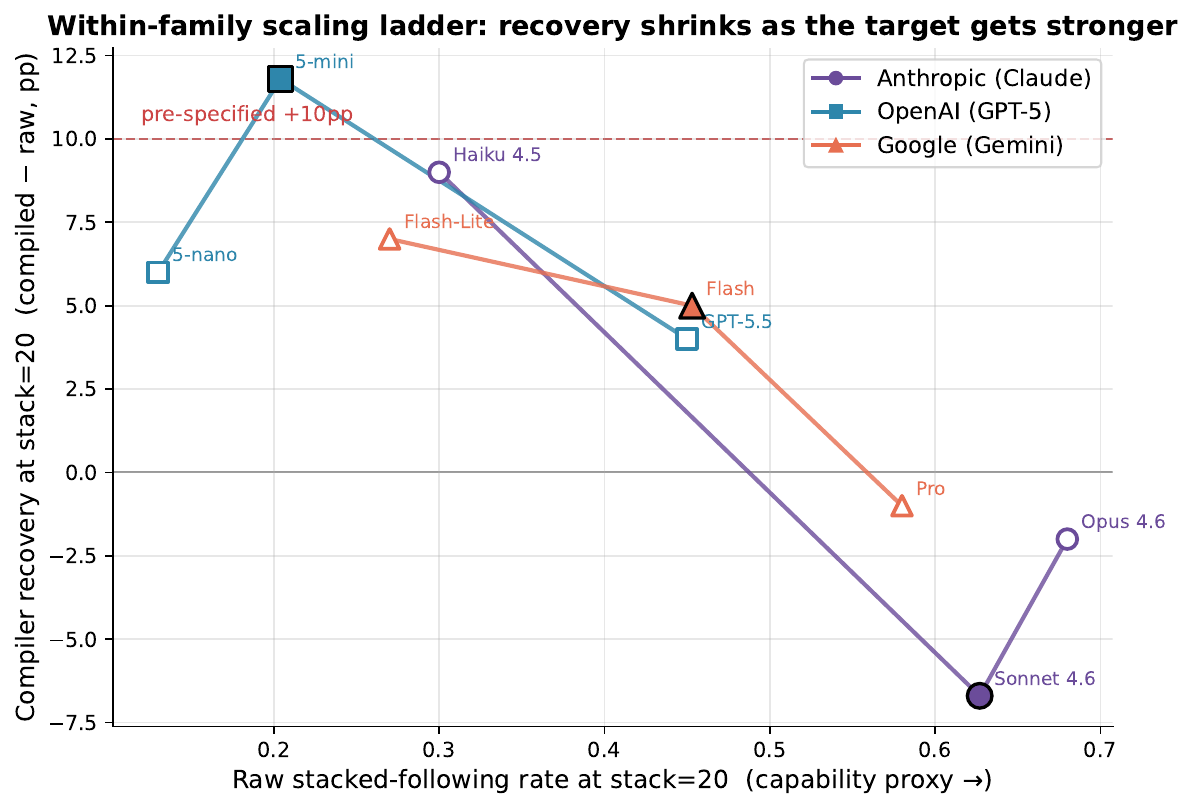}
\caption{Recovery at stack=20 vs.\ raw rate, one line per family. Filled markers are
the production-tier anchors; open markers the added rungs. Recovery falls as the
target strengthens; a floor effect appears at the very bottom.}
\label{fig:ladder}
\end{figure}

Holding the provider fixed and varying only size (Anthropic
Haiku$\to$Sonnet$\to$Opus; OpenAI nano$\to$mini$\to$5.5; Google
Flash-Lite$\to$Flash$\to$Pro), recovery at stack=20 falls as the target strengthens
(\cref{fig:ladder}, \cref{tab:ladder}). Across the nine targets,
\textbf{Spearman(raw rate, recovery) $= -0.85$ ($p=0.004$)}; the top rung of every
family gains the least (Opus $-2.0$, GPT-5.5 $+4.0$, Gemini-Pro $-1.0$~pp), and
recovery peaks for \emph{weak-but-not-floored} targets, with a floor effect at the
bottom (GPT-5-nano $+6.0 <$ GPT-5-mini $+11.8$). This is single-seed, stack=20 only,
and pooling families re-mixes the cross-lab confound, so we read it as corroboration,
not proof.

\begin{table}[t]
\centering
\caption{Ladder recovery (pp) at stack=20, compiled by Sonnet 4.6.}
\label{tab:ladder}
\begin{adjustbox}{max width=\columnwidth}
\begin{tabular}{lccc}
\toprule
Provider & small & mid (anchor) & large \\
\midrule
Anthropic & Haiku $+9.0$      & Sonnet $-6.7$ & Opus $-2.0$ \\
OpenAI    & nano $+6.0$       & mini $+11.8$  & 5.5 $+4.0$ \\
Google    & Flash-Lite $+7.0$ & Flash $+5.0$  & Pro $-1.0$ \\
\bottomrule
\end{tabular}
\end{adjustbox}
\end{table}

\subsection{Pre-registration scorecard}
\label{sec:scorecard}
We fixed three predictions before running the experiments, and report the outcome of
each. \textbf{(1)}~Decay is sigmoidal: \ding{51}{} for two of three models; Gemini is
non-sigmoidal. \textbf{(2)}~At least 15\% of pairs conflict: \ding{55}{} on the
satisfiable set (${\sim}12\%$; 17.7\% if the designed-in impossibilities are counted).
\textbf{(3)}~The compiler recovers $\ge$50\% of the loss with $d \ge 0.5$: \ding{55}{}
for every model (best $d=+0.43$). The contributions we claim are the predictions the
data confirmed, together with the capability-graded recovery that the third prediction
under-anticipated.

\section{Discussion}
\label{sec:discussion}

\textbf{Why does the benefit depend on capability?} Two readings are consistent with
the data, and they are not mutually exclusive. The \emph{practical} one is about cost:
the rewrite is computed once and reused, so it is an inexpensive intervention for the
small, fast models that serve most production traffic, and it is simply not needed for
top-tier models. The \emph{mechanistic} one is about what the rewrite supplies: a
strong model already recovers the stack's structure as it reads the raw prompt,
grouping the format rules and settling precedence on its own, whereas a weaker model
does not. Writing that structure out therefore helps the weaker model and adds little
for the stronger one. The conflict topology of \cref{sec:conflict} also explains why
the recovery is bounded: a precedence note cannot satisfy the 15 logically impossible
pairs, and only a verifier-guided compiler that dropped one side of each would close
that part of the gap.

\textbf{Limitations.} (i)~\emph{Construct:} verifiers capture the letter, not the
intent, of an instruction; ``target strength'' is a within-benchmark proxy, making
the strength--recovery link partly circular (broken by the cross-model and
same-baseline-control evidence). (ii)~\emph{Validation:} the verifier audit is
abbreviated (${\sim}500$ labels, 3 verifiers spot-checked only); absolute levels
carry small unaudited bias, but comparative claims cancel it. (iii)~\emph{Baseline:}
the compiler is only compared to machine-degraded versions of a naive dump, and we do
\textbf{not} compare against a competently hand-written prompt, and a single model
(Sonnet) does all compiling. (iv)~\emph{Scope:} compiler recovery is shown on advice
(and code as a robustness check); the ladder is single-seed at stack=20. None of the
headline claims extend beyond the tested grid.

\textbf{Future work.} The compiler studied here is deliberately minimal: it groups
and merges constraints, but it never removes one. Because 15 of the conflicting pairs
are logically unsatisfiable (\cref{sec:conflict}), no prompt-only rewrite can satisfy
both sides, which sets a ceiling on the achievable recovery. The natural next step is
a \emph{verifier-guided} compiler that detects such pairs, drops the lower-priority
instruction, and reports the omission to the caller, so that the residual conflict is
made explicit rather than silently violated. Two further directions would sharpen the
present results: a comparison against a competently hand-written prompt, rather than
only the machine-degraded baselines used here, and a multi-seed run of the
within-family ladder across all stack sizes.

\section{Conclusion}
\label{sec:conclusion}

Stacking many instructions into one prompt reliably degrades instruction-following.
The decline is non-linear, and it runs through an interpretable, reproducible set of
pairwise conflicts rather than uniform forgetting. A single training-free rewrite of
the prompt offers a targeted remedy: it returns a meaningful share of the lost follow
rate to weaker models (${+}11$~pp for the weakest), while strong models, which already
impose this structure themselves, are left essentially unchanged. The gain comes from
the rewrite rather than from extra tokens, reordering, or measurement headroom, and it
holds down three within-family scaling ladders. Because the rewrite is paid for once
and reused at no per-query cost, it is most worth applying precisely where
instruction-following is weakest.

%% file: appendix.tex
\section{Full Instruction Set}
\label{app:instructions}

The 24 atomic instructions; the 22 deterministic ones form the active pool. C2 and
C3 (starred) require an LLM judge and are excluded from all reported experiments.

\begin{itemize}[leftmargin=1.2em,itemsep=1pt]
\item \textbf{Format.} F1 valid JSON (\texttt{json.loads}, optional code-fence
stripped); F2 wrap whole response in a markdown code fence; F3 $\ge 3$ markdown
bullets; F4 trailing ``Summary: \ldots'' line on its own.
\item \textbf{Length.} L1 $\le 50$ words; L2 $\ge 150$ words; L3 exactly 5 bullets;
L4 title (first line) $\le 10$ words.
\item \textbf{Lexical.} X1 use ``therefore'' $\ge 1$; X2 avoid ``however''
entirely; X3 exactly three numbered citations \texttt{[1] [2] [3]}; X4 all
lowercase.
\item \textbf{Structural.} S1 begin with ``Answer:''; S2 include ``\#\# Pros'' and
``\#\# Cons''; S3 second-person $\ge 3$; S4 no question marks.
\item \textbf{Reasoning.} R1 ``\#\# Reasoning'' header before the answer; R2 ``\#\#
Assumptions'' ($\ge 2$ items); R3 ``Confidence: 0.XX'' line; R4 ``\#\# Alternatives''
($\ge 2$ items).
\item \textbf{Content.} C1 reference a year 1900--2025; C2$^*$ exactly one
analogy/metaphor; C3$^*$ a counterargument; C4 $\ge 2$ named sources.
\end{itemize}

\section{Extended Related-Work Comparison}
\label{app:related}

\begin{table}[h]
\centering
\caption{Full benchmark comparison.}
\begin{adjustbox}{max width=\columnwidth}
\begin{tabular}{lccccc}
\toprule
Property & IFEval & FollowBench & RECAST & MulDimIF & \textbf{Ours} \\
\midrule
Instructions/prompt & 1--2 & incr., 1 cat & 30+ & multi & \textbf{1--20, 6 cat} \\
Cross-category stacking & no & no & partial & partial & \textbf{yes (uniform)} \\
Deterministic verifiers & yes & partial & partial & partial & \textbf{22/24} \\
Pairwise conflict analysis & no & no & data & data & \textbf{231 pairs, FDR} \\
Impossible-vs-behavioral & no & no & no & no & \textbf{yes} \\
Primary purpose & eval & eval & train & train & \textbf{eval + mitig.} \\
Training-free mitigation & no & no & no & no & \textbf{yes (compiler)} \\
\bottomrule
\end{tabular}
\end{adjustbox}
\end{table}

\section{Verifier Validation}
\label{app:verifiers}

Each verifier was audited against hand labels formed blind to its verdict. After an
L4 bug fix (a leading code-fence line was wrongly counted as a short title; corrected
and re-scored from cache), 18 of 19 audited verifiers match humans exactly; C4 is
over-strict by three false negatives in fifteen. R1, R3, S2 were spot-checked only; a
full 100-label-per-instruction audit is owed.

\begin{table}[h]
\centering
\caption{Verifier-vs-human agreement.}
\begin{adjustbox}{max width=\columnwidth}
\begin{tabular}{lcc}
\toprule
Verifier & $n$/sheet & agreement \\
\midrule
F1--F4, L1--L4, X1--X4, S1, S3, S4, R2, R4, C1 (18) & 15--31 & 1.000 \\
C4 ($\ge 2$ named sources) & 15 & 0.800 \\
\bottomrule
\end{tabular}
\end{adjustbox}
\end{table}

\section{Category and Per-Instruction Degradation}
\label{app:categories}

\begin{figure}[h]
\centering
\includegraphics[width=0.8\columnwidth]{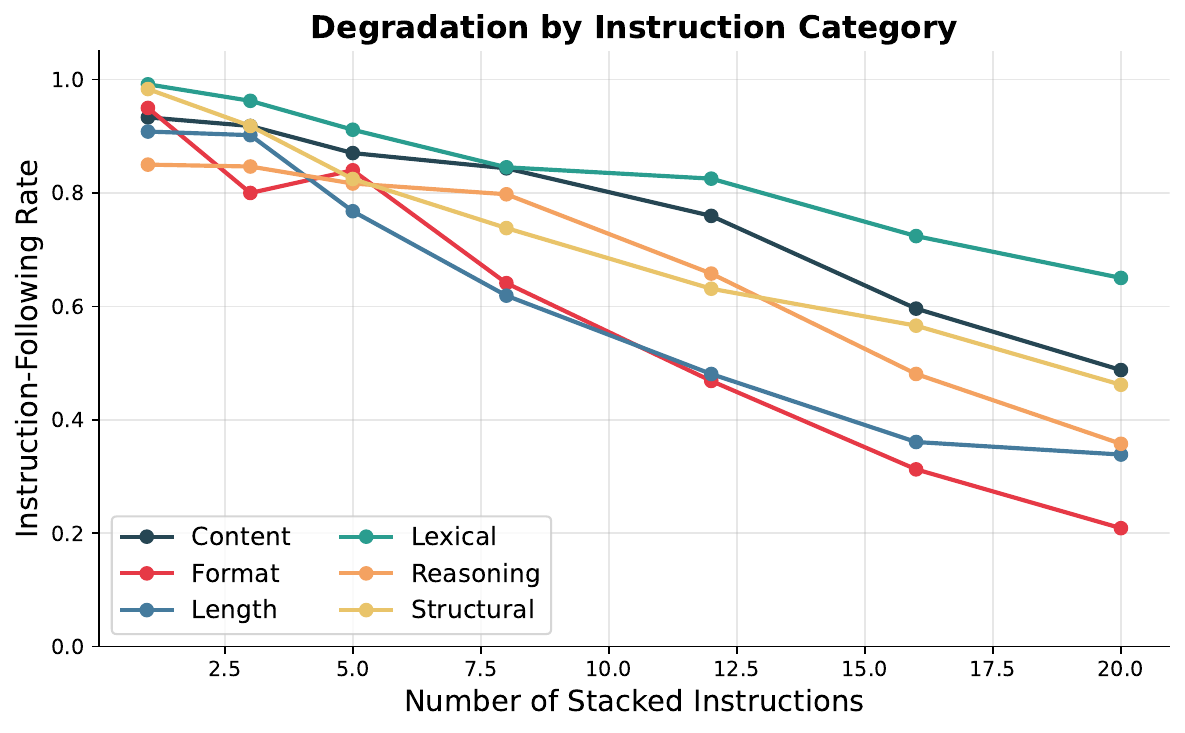}
\caption{Mean follow rate by category, pooled across models. Lexical survives; format
and length collapse hardest.}
\end{figure}

\begin{figure}[h]
\centering
\includegraphics[width=\columnwidth]{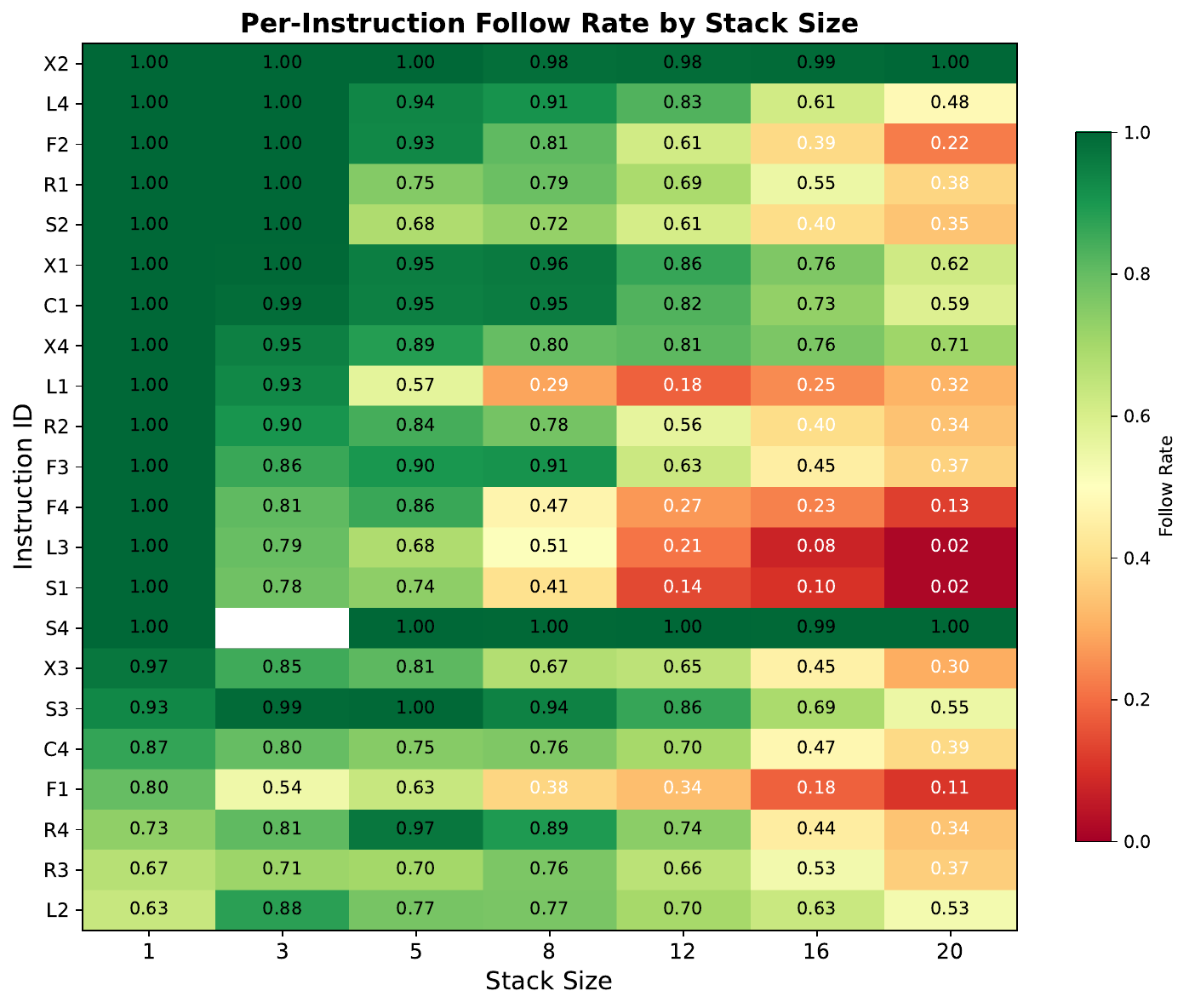}
\caption{Per-instruction follow rate by stack size. L3, S1, F4, F1 fall fastest; S4,
X2, X4 stay high.}
\end{figure}

\section{Conflict Topology: Details}
\label{app:conflict}

\textbf{The 15 logical impossibilities.} F1 (JSON) $\times$ \{F3,F4,L3,R1,R2,R3,R4,
S1,S2\}; X4 (lowercase) $\times$ \{S1,F4,R3\}; F2 (code fence) $\times$ \{S1,F4\};
and L1 $\times$ L2.

\begin{table}[h]
\centering
\caption{Ten strongest \emph{behavioral} conflicts on Sonnet 4.6 (impossibilities
excluded), sorted by observed $-$ expected.}
\begin{adjustbox}{max width=\columnwidth}
\begin{tabular}{lccccc}
\toprule
pair & P(A) & P(B) & obs & exp & interaction \\
\midrule
F1--L4 & 1.00 & 1.00 & 0.00 & 1.00 & $-1.00$ \\
L3--R2 & 1.00 & 1.00 & 0.00 & 1.00 & $-1.00$ \\
F4--R3 & 1.00 & 0.95 & 0.00 & 0.95 & $-0.95$ \\
F1--X3 & 1.00 & 0.90 & 0.10 & 0.90 & $-0.80$ \\
L3--R4 & 1.00 & 0.75 & 0.00 & 0.75 & $-0.75$ \\
X3--C4 & 0.90 & 0.85 & 0.05 & 0.77 & $-0.71$ \\
F2--R4 & 1.00 & 0.75 & 0.05 & 0.75 & $-0.70$ \\
X3--R1 & 0.90 & 1.00 & 0.30 & 0.90 & $-0.60$ \\
L1--R1 & 1.00 & 1.00 & 0.40 & 1.00 & $-0.60$ \\
F3--X3 & 1.00 & 0.90 & 0.45 & 0.90 & $-0.45$ \\
\bottomrule
\end{tabular}
\end{adjustbox}
\end{table}

\begin{table}[h]
\centering
\caption{Significant pairwise interactions per model (advice, 231 pairs, FDR 0.05).
GPT-5-mini has the \emph{fewest} conflicts despite collapsing hardest---accumulation,
not a denser conflict set, drives its aggregate collapse.}
\begin{adjustbox}{max width=\columnwidth}
\begin{tabular}{lcccc}
\toprule
Model & Conflicts & impossib. & behav. & Synergies \\
\midrule
Sonnet 4.6 & 41 & 15 & 26 & 42 \\
Gemini 2.5 Flash & 48 & 14 & 34 & 48 \\
GPT-5-mini & 36 & 13 & 23 & 27 \\
\bottomrule
\end{tabular}
\end{adjustbox}
\end{table}

\begin{figure}[h]
\centering
\includegraphics[width=\columnwidth]{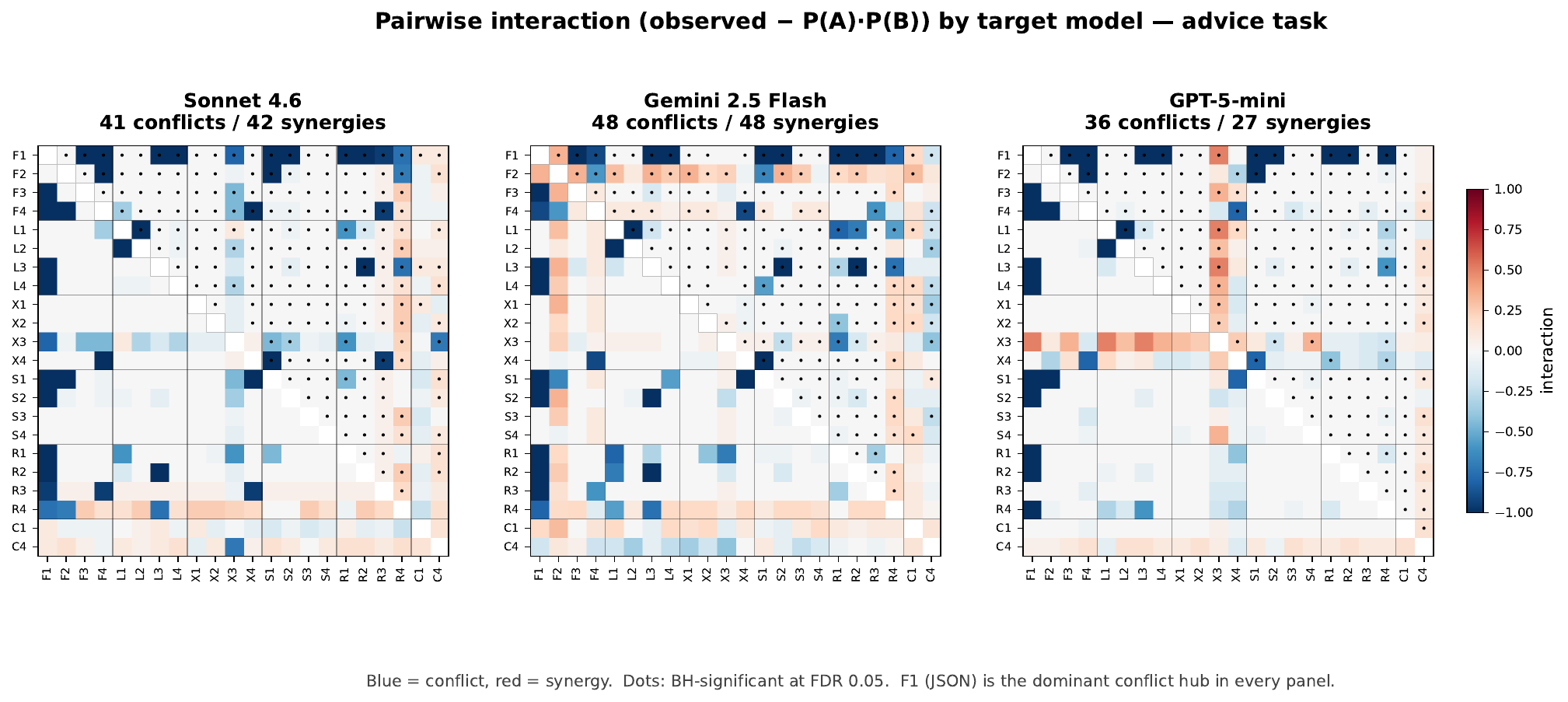}
\caption{Per-model conflict heatmaps. F1 is the dominant conflict hub in every panel;
GPT-5-mini (right) is the sparsest despite collapsing hardest in aggregate.}
\end{figure}

\section{Multi-Task Degradation}
\label{app:multitask}

\begin{table}[h]
\centering
\caption{Mean follow rate (95\% CI), GSM8K math.}
\begin{adjustbox}{max width=\columnwidth}
\begin{tabular}{lcccc}
\toprule
Model & stack=1 & stack=5 & stack=16 & stack=20 \\
\midrule
Sonnet 4.6 & 0.955 & 0.853 & 0.606 & 0.591 \\
GPT-5-mini & 0.936 & 0.841 & 0.323 & 0.206 \\
Gemini 2.5 F & 0.918 & 0.814 & 0.557 & 0.431 \\
\bottomrule
\end{tabular}
\end{adjustbox}
\end{table}

\begin{table}[h]
\centering
\caption{Mean follow rate (95\% CI), HumanEval code. Per-model ordering flips:
GPT-5-mini ties Sonnet; Gemini is weakest.}
\begin{adjustbox}{max width=\columnwidth}
\begin{tabular}{lcccc}
\toprule
Model & stack=1 & stack=5 & stack=16 & stack=20 \\
\midrule
Sonnet 4.6 & 0.964 & 0.837 & 0.577 & 0.447 \\
GPT-5-mini & 0.914 & 0.793 & 0.521 & 0.443 \\
Gemini 2.5 F & 0.841 & 0.777 & 0.487 & 0.344 \\
\bottomrule
\end{tabular}
\end{adjustbox}
\end{table}

\begin{figure}[h]
\centering
\includegraphics[width=\columnwidth]{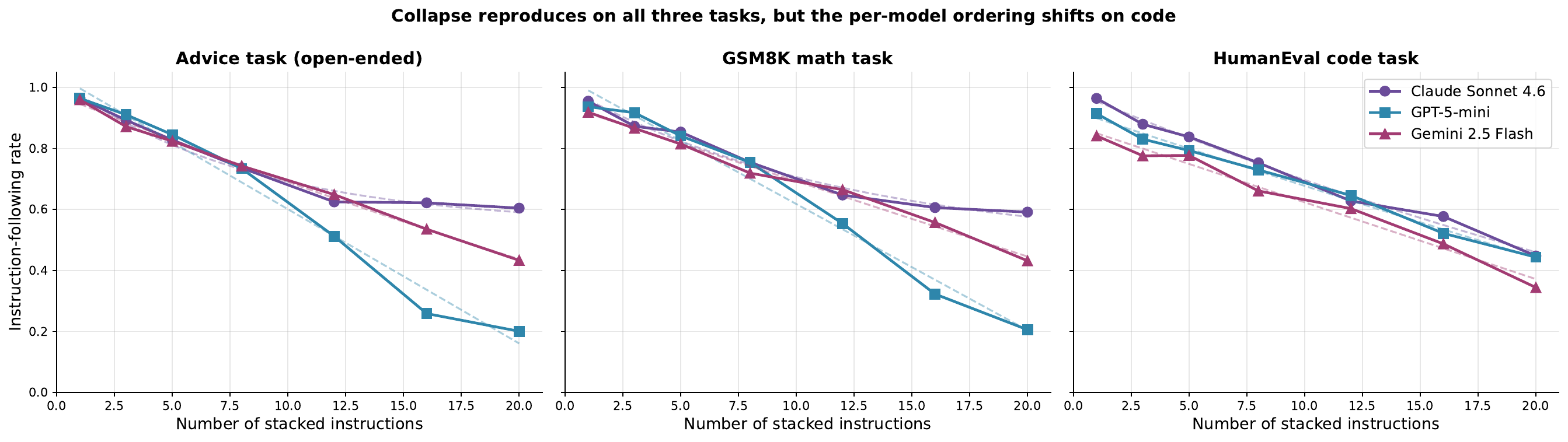}
\caption{Degradation on advice / math / code. The \emph{fact} of collapse
generalizes; the per-model \emph{ordering} does not.}
\end{figure}

\section{Compiler: Conditions and Ablations}
\label{app:compiler}

\begin{figure}[h]
\centering
\includegraphics[width=\columnwidth]{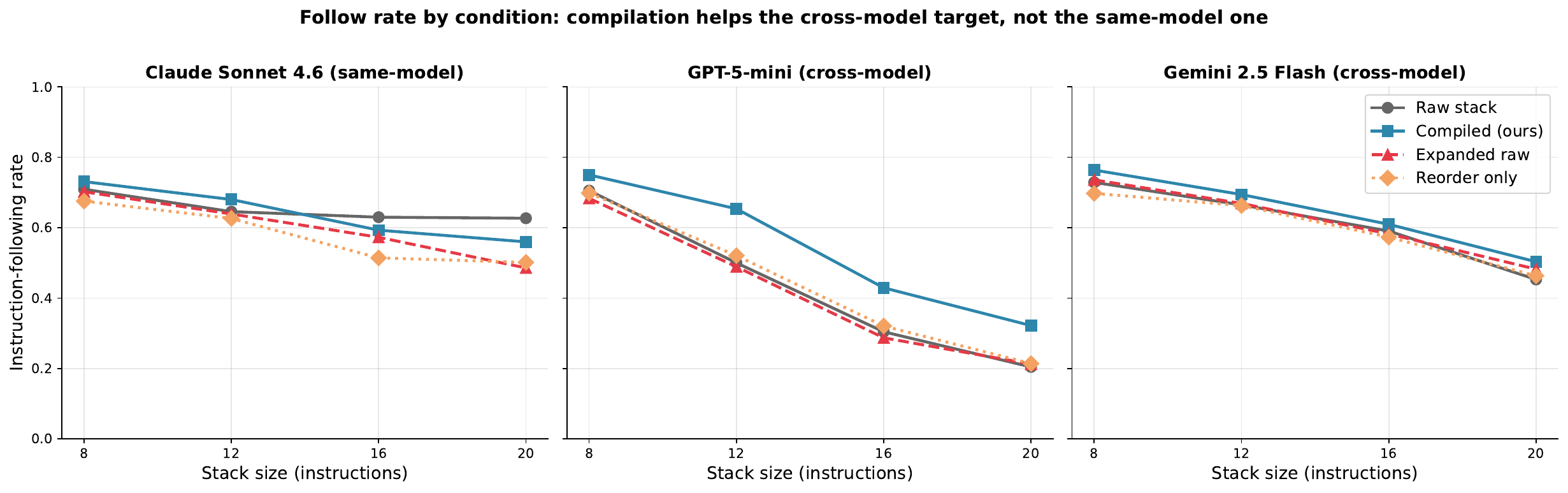}
\caption{Follow rate by condition $\times$ stack size per target. \texttt{compiled}
separates from controls on the cross-model targets; on Sonnet \texttt{raw} is best.}
\end{figure}

\begin{figure}[h]
\centering
\includegraphics[width=\columnwidth]{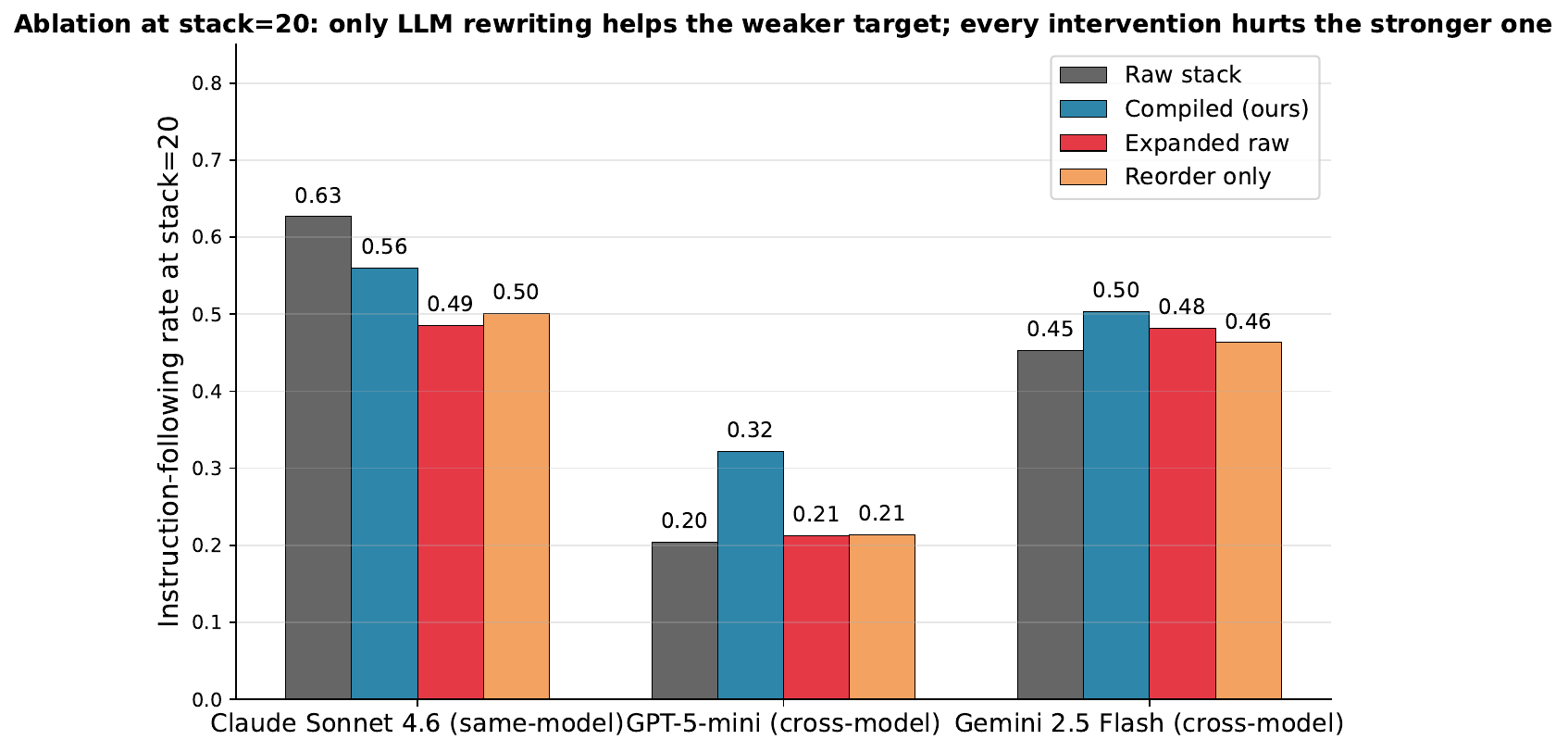}
\caption{Stack=20 conditions per target. Only the LLM rewrite (\texttt{compiled})
separates from \texttt{raw}; token-padding and reordering do not.}
\end{figure}

\section{Statistical Methods}
\label{app:stats}

All inference is reproduced by \texttt{scripts/extra\_stats.py} on the released
parquets (numpy/scipy only).

\textbf{Cluster bootstrap.} Because the 15 items under one stack share a single
compiled prompt and instruction draw, items are not independent. We resample whole
stacks with replacement (3{,}000 reps) and recompute the statistic, giving CIs and
two-sided $p$-values that respect within-stack correlation. The naive
$n{=}1{,}200$-paired-observation t-test overstates significance; under the cluster
bootstrap the two cross-model recoveries survive ($p\le0.001$) while the
same-model ``harm'' does not ($p=0.49$ pooled; $p=0.037$ at stack=20 only).

\textbf{Model-selection bootstrap.} For each degradation curve we resample responses
within each stack size, recompute the mean curve, refit linear/exponential/sigmoid,
and record the AIC winner (400 reps). Sigmoid wins 93\% (Sonnet) and 100\%
(GPT-5-mini) of refits; Gemini selects sigmoid in $<2\%$ (linear 64\%, exp 35\%).

\textbf{Interaction inference.} For each pair, interaction $=
\text{observed}(A\wedge B) - P(A)\,P(B)$ with $P(\cdot)$ from singleton baselines;
significance by Benjamini--Hochberg at FDR 0.05. The feasible-only conflict rate is
${\sim}12\%$ with a pair-resampling 95\% CI of $[4.6, 11.6]$. Cross-model Spearman of
all 231 interactions: $+0.27$, $+0.28$, $+0.23$ (all $p<10^{-3}$).

\textbf{Ladder.} Spearman between raw stacked-following rate and recovery across the
nine ladder targets is $-0.85$ ($p=0.004$, $n=9$, single seed).